\documentclass[12pt]{article}

\newcommand{\beq}{\begin{eqnarray}}
 \newcommand{\eeq}{\end{eqnarray}}
\newcommand{\be}{\begin{equation}}
 \newcommand{\ee}{\end{equation}}

\def\fun#1#2{\lower3.6pt\vbox{\baselineskip0pt\lineskip.9pt
\ialign{$\mathsurround=0pt#1\hfil ##\hfil$\crcr#2\crcr\sim\crcr}}}

\newcommand{{\SD}}{\rm SD}

\newcommand{\vex}{\mbox{\boldmath${\rm x}$}}
\newcommand{\vew}{\mbox{\boldmath${\rm w}$}}
\newcommand{\vey}{\mbox{\boldmath${\rm y}$}}
\newcommand{\ver}{\mbox{\boldmath${\rm r}$}}

\newcommand{\vez}{\mbox{\boldmath${\rm z}$}}

\newcommand{\lan}{\langle}
\newcommand{\ran}{\rangle}

\title{Confinement of light quarks in QCD at finite density.
}
\author{ Yu.A. Simonov\\
State Research
Center\\Institute of Theoretical and Experimental Physics, \\
Moscow, 117218 Russia}

\begin{document}
\maketitle

\begin{abstract}

 The $4q$ effective Lagrangian and the gap equation are derived
 for light quarks in the confinement phase of QCD. The
 modification of the confining string  due to finite quark density
 (chemical quark potential $\mu$) is observed.
As a surprising result in a multiquark system with a common string junction
an attractive well  appears of radius $\mu/\sigma$ and of an average depth equal to $\mu$. Possible
implications for the density phase transition are discussed.

\end{abstract}

\section{Introduction}

Recently the high-density effects in QCD attracted a lot of
attention  because  of the possible density
 phase transition \cite{2}, and interesting accompanying
 phenomena like Color Superconductivity (CS)  \cite{3} (for earlier
 papers see \cite{4}). On the fundamental level the physical expectation of any  phase
 transition  may  be connected to the possible reconstruction
 of the vacuum  and  for that one needs that the energy density to be
  of the order of the vacuum energy  density $\varepsilon_{cr}\sim \varepsilon_{vac} \approx
-\frac{11}{3} N_c\frac{\alpha_s}{32\pi} \lan (F^a_{\mu\nu}
 )^2\ran, \varepsilon_{cr} \sim 1$ GeV/fm$^3$ which provides the
 drastic change in the vacuum structure and may cause the phase
 transition. In case of temperature phase transition the
 corresponding energy density $\varepsilon$ is indeed of the
 order of $\varepsilon_{cr}$,  and as it was first argued
 in \cite{6}  and later
 measured on the  lattice \cite{7}, the QCD vacuum is strongly
 transformed in a way, that most part of  colorelectric fields
 evaporate above $T_c$. In this way
 $T_c$ was calculated through $\varepsilon_{cr}$ and  found in good agreement with lattice data \cite{6}.
Also, it was shown that in QCD in the framework of hadron resonance gas
and making use of the low-energy theorems at $T \neq 0$ \cite{letqcd}
the quark condensate and one half (colorelectric component)
of gluon condensate evaporate at the same temperature \cite{agfed}, which corresponds to the temperature of
quark-hadron phase transition. Besides application of the effective dilaton Lagrangian to gluodynamics
\cite{dilaton} and to QCD with light and heavy hadrons \cite{agebilg} permitted to conclude that similar
reconstruction of the nonperturbative gluon vacuum takes place at finite temperature.

  A similar arguments for the density phase
 transition would imply that at the baryon density of $\sim 1$
nucleon/fm$^3$, i.e. $3 \div 6$ times higher than the standard nuclear
density, the vacuum can be reconstructed in such a way, that part
of fields, e.g. the colorelectric fields responsible for
confinement, disappear above critical density.

A completely different route was trodden  by the groups who
studied the  phenomenon of color superconductivity \cite{3}. Here
main argument was the model study of CS in the formalism of the
NJL or instanton model (see \cite{8} for a recent review) and  the
practically important question of coexistence of quark and nuclear
matter  was also studied   lately in the unified NJL approach
\cite{9}. In this strategy the question  of the vacuum
reconstruction is not addressed, however it is usually assumed
that confinement disappears  in the course of establishing of the
CS dynamics. On the other hand the possible importance of vacuum
fields for the density transition  and CS was suggested in
\cite{10}, but no detailed theory was presented there. It is the
purpose of the present paper to start the investigation of the
role of density on the vacuum fields in general and confinement in
particular.

In this  paper we ask ourselves a short and simple question: how
the nonzero baryonic chemical potential $\mu$ acts on the
confinement of light quarks, and come  to the  unexpected answer,
that the  confining string of light quarks is destroyed gradually
by $\mu$ in such a way, that the one part of string,  near the
string junction is eaten by the nonzero $\mu $, while at distant
$r$  the string survives.
The plan of the paper is as follows. In section 2 the
effective Lagrangian is derived from the QCD Lagrangian with
nonzero $\mu$ and their solution is discussed in section 3.
 Section 4 is devoted to the physical implications of results and prospectives.

\section{Derivation of Effective $4q$ Lagrangian}

  One starts with the QCD partition function  in
   presence of quark chemical potential $\mu$ in the Euclidean
  space-time, and we begin with the zero temperature, $T=0$.

  \be
Z=\int DAD\psi D\psi^+ e^{-S_0(A)+\int~^f\psi^+(i\hat
\partial+im- i\mu\gamma_4+g\hat A)~^f\psi d^4x}
\label{1}\ee where $S_0(A)=\frac{1}{4}\int(F^a_{\mu\nu}(x))^2
d^4x$, $m$ is the current quark mass (mass matrix $\hat m$ in
SU(3)), and the quark operator $^f\psi_{a\alpha}(x)$ has flavor
index $a(f=1,... n_f)$, color index $a(a=1, ... N_c)$ and Lorenz
bispinor index $\alpha(\alpha=1,2,3,4)$, and  we use the contour
gauge \cite{11} to express $A_\mu(x)$ in terms of $F_{\mu\nu}$.
 One has for the contour $z_\mu(s,x)$ starting at point $x$ and ending at
$Y=z(0,x)$
 \be A_\mu(x)=\int^1_0 ds\frac{\partial
z_\nu(s,x)}{\partial s}\frac{\partial z_\rho(s,x)}{\partial x_\mu}
F_{\nu\rho} (z(s))\equiv \int^x_Y d \Gamma_{\mu\nu\rho} (z)
F_{\nu\rho} (z) .
\label{2} \ee
 Integrating out the gluonic fields
$A_{\mu} (x)$, one obtains
\be
Z=\int D\psi D\psi^+
e^{\int~^f\psi^+(i\hat \partial+im- i\mu\gamma_4)^f\psi d^4x}
e^{L^{(2)}_{EQL}+L^{(3)}_{EQL}+...}
\label{3} \ee
 where the EQL
proportional to $\lan\lan A^n\ran\ran$ is denoted by
$L_{EQL}^{(n)}$,
\be L^{(2)}_{EQL}=\frac{g^2}{2}\int
d^4xd^4y~^f\psi^+_{a\alpha}(x)
~^f\psi_{b\beta}(x)~^g\psi^+_{c\gamma}(y)~^g\psi_{d\varepsilon}(y)
\lan  A^{(\mu)}_{ab}(x) A^{(\nu)}_{cd}(y)\ran
\gamma^{(\mu)}_{\alpha\beta} \gamma^{(\nu)}_{\gamma\varepsilon}
\label{4} \ee
Average of gluonic fields can be computed using
(\ref{2}) as (see \cite{12} for  details of derivation)
\be g^2\lan A^{(\mu)}_{ab}(x) A^{(\nu)}_{cd}(y)\ran=
\frac{\delta_{bc}\delta_{ad}}{N_c} \int^x_0
du_i\alpha_{\mu}(u)\int^y_0 dv_k\alpha_\nu(v)
D^{(E,H)}(u-v)(\delta_{\mu\nu}\delta_{ik}-\delta_{i\nu}\delta_{k\mu}),
\label{5} \ee
 where $D^{(E,H)}(x)$ is the correlator $\lan E_i(x)
E_i(0)\ran$ or $\lan H_i(x) H_i(0)\ran$. As it was argued in
\cite{12} the dominant contribution at large distances from
 the static antiquark is given by the color-electric fields,
 therefore at the first stage  we shall write down
 explicitly $L_{EQL}^{(2)} (el)$ for this case, i.e. taking $\mu=\nu=4$. As a result one
 has\cite{12}
\be
 L_{EQL}^{(2)}(el)=\frac{1}{2N_c}\int d^4x\int
d^4y~^f\psi^+_{a\alpha}(x)~^f\psi_{b\beta}(x)
~^g\psi^+_{b\gamma}(y)~^g\psi_{a\varepsilon}(y)
\gamma^{(4)}_{\alpha\beta}\gamma^{(4)}_{\gamma\varepsilon} J^E(x,y)
\label{6} \ee
where $J^E(x,y)$ is
\be
 J^E(x,y) =\int^x_0 du_i\int^y_0
dv_i D^E(u-v),~~i=1,2,3.
 \label{7} \ee
 One can form bilinears
$\Psi^{fg}_{\alpha\varepsilon}\equiv
~^f\psi^+_{a\alpha}~^g\psi_{a\varepsilon}$ and project using Fierz
procedure given isospin and Lorentz structures,
$\Psi^{fg}_{\alpha\varepsilon}\to \Psi^{(n,k)}(x,y).$ Here we
consider only $\psi^+\psi$ bosonization.
 With the
help of the standard bosonization trick (here $\tilde J \equiv
\frac{1}{N_c} J^E$)
\be e^{-\Psi\tilde J\Psi}= \int(\det \tilde
J)^{1/2} D\chi\exp [-\chi \tilde J\chi+ i\Psi\tilde J \chi + i\chi
\tilde J \Psi] \label{8} \ee \be Z=\int D\psi D\psi^+ D\chi \exp
L_{QML} \label{9} \ee
 one obtains the effective Quark-Meson Lagrangian (QML)
$$ L^{(2)}_{QML} =\int d^4x\int
d^4y\left\{~^f\psi^+_{a\alpha}(x)[(i\hat\partial+im-i\mu\gamma_4)_{\alpha\beta}\delta(x-y)
+iM^{(fg)}_{\alpha\beta} (x,y)]~^g\psi_{a\beta}(y)- \right.$$ \be
\left.-\frac{1}{N_c}\chi^{(n,k)}(x,y) J^E(x,y)
\chi^{(n,k)}(y,x)\right\} \label{10} \ee
 and the effective  quark-mass operator is
 \be
M^{(fg)}_{\alpha\beta}(x,y) =\sum_{n,k} \chi^{(n,k)}(x,y)
O^{(k)}_{\alpha\beta}t^{(n)}_{fg}\tilde J(x,y).
\label{11}
\ee
Here the operator $\hat{O}$ is a set of all irreducible combinations of Dirac matrixes.

The QML in Eq.(\ref{10}) $L^{(2)}_{QML}$ contains  functions $
\chi^{(n,k)}$  which are integrated out in (\ref{9}), and the
standard way is to find $\chi^{(n,k)}$ from the  stationary point
of $L^{(2)}_{QML}$. Limiting oneself to the scalar and
pseudoscalar fields and using the nonlinear parametrization one
can write for the operator $\hat M$ in (\ref{10})
 \be \hat
M(x,y)=M_{S}(x,y) \hat U(x,y),\hat U=exp(i\gamma_{5}\hat\phi),
\hat \phi(x,y)=\phi^{f}(x,y) t^{f}. \label{12}\ee After
integrating out the quark fields one obtains the ECL in the form
\be L^{(2)}_{ECL}(M_S,\hat \phi)=-2n_f N_c( J^E (x,y))^{-1}
M^2_S(x,y)+ N_c tr\log[(i\hat\partial+im-i\mu\gamma_4)\hat 1+iM_S
\hat U]. \label{13}\ee
 The stationary point equations $\frac{\delta
L^{(2)}_{ECL}}{\delta M_s}= \frac{\delta L^{(2)}_{ECL}}{\delta
\hat \phi}=0$ at $\hat \phi=\hat\phi_0$, $M_s=M_s^{(0)}$
immediately show that $\hat \phi_0=0$ and $M^{(0)}_s$ satisfies
nonlinear equation
\be
 i M^{(0)}_{S}(x,y)={4} tr S J^E(x,y)=
 (\gamma_4 S \gamma_4)  J^E(x,y) ,~~
 S(x,y)=-[i\hat\partial+im -i\mu\gamma_4+iM_S \hat U]^{-1}_{x,y}.
 \label{14}\ee

This equation plays the role of the gap equation and
 is the main point of our further investigation. For $\mu=0$ this was done in \cite{12}
 and in the next section we find how results of \cite{12} are modified by the nonzero $\mu$.

 \section{The confining string at nonzero $\mu$}

Our  basic  equations (\ref{6}),(\ref{7}) are nonlocal in time because of the
integral over $dx_4 dy_4$ in (10). This nonlocality and the parameter
which it governs can be handled most  easily, when one uses instead
of $M(z,z')$, $S(z,z')$ the  Fourier  transforms.

\be
S(z_4-z'_4,\vez,\vez')= \int e^{ip_4(z_4-z'_4)}
S(p_4,\vez, \vez') \frac{dp_4}{2\pi}\label{15}
\ee
and the same for $M(z,z')$. Then from (\ref{14}) one obtains  a
system of equations
\be
(\hat p_4-i\hat \partial_z-im+ i\mu\gamma_4)S(p_4, \vez, \vew) - i\int
M(p_4,\vez, \vez') S(p_4, \vez', \vew) d \vez'=
\delta^{(3)}(\vez - \vew)\label{16}
\ee

To simplify matter, one assumes for $D^E(x)$ the Gaussian form,
$D^E(x)=D(0) \exp \left( -\frac{x^2}{4T^2_g}\right)$. Then  for
$M(p_4, \vez, \vew)$ one has
$$
iM(p_4, \vez, \vew)= 2\sqrt{\pi}T_g \int \frac{dp'_4}{2\pi}
e^{-(p_4-p'_4)^2 T_g^2}\times
 $$
 \be
 \times [J^E(\vez, \vew)\gamma_4S(p'_4,\vez,
 \vew)\gamma_4]\label{17}
 \ee
 where  $J^E$
 is defined in (\ref{7}) and we have factored out the time--dependent
 exponent, using the  Gaussian  representation of $D(u)$.

 All dependence of $M$ on $p_4$ as can be seen in (\ref{17}) is due to the
 factor $\exp [-(p_4-p'_4)^2T_g^2]$    and disappears in the limit when
 $T_g$ goes to zero, while the string tension $\sigma\sim D(0)T^2_g$
 is kept  fixed. This limit can be called the string limit of QCD,
 and we shall study its consequences for equations (\ref{16}),(\ref{17}) in this
 section.

 So in the string limit, with $M$ independent of $p_4$, let us
 consider the Hermitian Hamiltonian
 \be
 \hat H\psi_n \equiv (\frac
 {\alpha_i}{i}\frac{\partial}{\partial z_i}+\beta
 m-\mu)\psi_n(\vez) + \beta \int M (p_4=0, \vez, \vez')\psi_n(\vez') d^3 \vez'= \tilde \varepsilon_n (\mu)\psi_n (\vez)\label{18}
 \ee
with eigenfunctions $\psi_n$ satisfying usual orthonormality
condition
$$
\int \psi^+_n(x)\psi_m( x) d^3 x=\delta_{nm},
$$
From (\ref{18}) it is clear, that one can redefine
 $\tilde\varepsilon_n(\mu) +\mu\equiv\varepsilon_n$, and $\varepsilon_n$ and $\psi_n$ do
not  depend  on $\mu$. Therefore in all subsequent formulas one
can use the same equations as in \cite{12}, but with the
replacement $\varepsilon_n \to \varepsilon_n - \mu$. In
particular, the Green's function $S$ can be expressed as \be
S(p_4,\vex,
\vey)=\sum_n\frac{\psi_n(\vex)\psi^+_n(\vey)}{p_4\gamma_4-i(\varepsilon_n-\mu)\gamma_4}\label{19}
\ee Inserting (\ref{19}) into (\ref{17}) one  has integrals of
the type: \be
\int^{\infty}_{-\infty}\frac{dp'_4}{2\pi}\frac{e^{-(p_4-p'_4)^2T^2_g}}
{(p'_4 \gamma_4-i(\varepsilon_n-\mu)\gamma_4)}=\frac{i}{2}\gamma_4
sign(\varepsilon_n-\mu)(1+0(p_4T_g,|\varepsilon_n|T_g)\label{20}
\ee
 Note, however, that the result depends on the boundary
conditions. If, e.g., one imposes the causality--type boundary
condition, then one obtains
$$
\int\frac{dp'_4}{2\pi}\frac{e^{ip'_4h_4}}{\gamma_4(p'_4-i(\varepsilon_n-\mu))}=
\left\{
\begin{array}{ll}
i\gamma_4e^{-\varepsilon h_4}\theta(\varepsilon_n-\mu),& h_4>0\\
-i\gamma_4\theta(\mu-\varepsilon_n)e^{\varepsilon h_4},& h_4<0
\end{array}
\right.\label{21}
$$

We are  thus led  to the following expression for $M$ in the
string limit \be M(p_4=0, \vez, \vew)= \sqrt{\pi}
T_g J^E(\vez,\vew)\gamma_{4}\Lambda(\vez, \vew)
\label{22}
\ee
where the definition is used
\be
\Lambda(\vez, \vew) = \sum_n\psi_n(\vez) sign (\varepsilon_n-\mu)
\psi^+_n(\vew)\label{23}
\ee

Let us  disregard for the moment the possible appearance in $M$ of
the vector component (proportional to $\gamma_{\mu},\mu=1,2,3,4)$
and concentrate on the scalar contribution only, since that is
responsible for CSB and confinement. Then one can look for
solutions of the Dirac equation (\ref{18}) in the following form
\cite{12} \be \psi_n(\vec r)=\frac{1}{r}\left (
\begin{array}{l}
G_n(r)\Omega_{jlM}\\
iF_n(r)\Omega_{jl'M}
\end{array}
\right)\label{24} \ee where $l'=2j-l$, and introducing the
parameter $\kappa(j,l)=(j+\frac{1}{2}) sign (l-j)$, and replacing
$M$ by a local operator (the generalization to the nonlocal case
is straightforward but cumbersome, for a possible change in the  nonlocal case see \cite{12}).
 We obtain a system of equations \be
\left\{
\begin{array}{l}
\frac{dG_n}{dr}+\frac{\kappa}{r}G_n-(\varepsilon_n-\mu+m+M(r))F_n=0\\
\frac{dF_n}{dr}-\frac{\kappa}{r}F_n+(\varepsilon_n-\mu-m-M(r))G_n=0\\
\end{array}
\right.\label{25}
\ee
Eq.(\ref{24}) possesses a symmetry $(\varepsilon_n-\mu, G_n,
F_n,\kappa)\leftrightarrow
(\mu-\varepsilon_n,
F_n,G_n,-\kappa)$
which means that for any solution of the form (\ref{23}) corresponding to
the eigenvalue $\varepsilon_n-\mu$, there is another solution of the form
\be
\psi_{\mu-\varepsilon_n}(r)=\frac{1}{r}\left (
\begin{array}{l}
F_n(r)\Omega_{jl'M}\\
iG_n(r)\Omega_{jlM}
\end{array}
\right )\label{26}
\ee
corresponding to the eigenvalue $(\mu-\varepsilon_n)$.

Therefore the difference, which enters (\ref{23}) can computed in
terms of $F_n,G_n$ as follows \be \Lambda(\vez, \vew) = \Lambda_0
(\vez, \vew) -\Delta \Lambda (\vez, \vew)\label{26a}\ee where
$\Lambda_0$ is the value of $\Lambda$ for $\mu=0$, i.e.  the same
as in  \cite{12}, while $\Delta \Lambda$ is defined as \be \Delta
\Lambda(\vez, \vew)=  2\sum_{0<\varepsilon_n<\mu}\psi_n(\vez)
\psi^+_n(\vew).\label{27}\ee Using decomposition (\ref{24}) one
can write $\Delta \Lambda$ as \be \Delta \Lambda (\ver, \ver')
=2\sum_{0<\varepsilon_n ,\mu} \left( \begin{array} {ll}
G_nG_n^+ \Omega \Omega^+,&-iG_nF_n^+ \Omega \Omega^{'+}\\
i F_nG_n^+ \Omega' \Omega^+,&F_nF_n^+ \Omega' \Omega^{'+}\end{array}\right) \label{28}\ee
where we have denoted $\Omega\equiv \Omega_{jlM}, \Omega' =\Omega_{jl'M}$,
 and we disregard nondiagonal part of $\Lambda$.

At this point one can follow the relativistic WKB method for Dirac
equation \cite{13} applied
 to calculation of $\Lambda$ in \cite{12} in case of $\mu=0$.
The classically available region for $\psi_n (\ver) $ with energy
$\varepsilon_n \equiv \varepsilon$ is $r_{\min}\leq r \leq
r_{\max}$, where $r_{\max, \min} = \frac{\varepsilon^2 \pm
\sqrt{\varepsilon - 4\sigma^2\kappa^2}}{2\sigma^2}$, and the
summation over $n$ in (\ref{27}) transforms into integration over
$d\varepsilon$, with the lower limit  (for a given $r$)
$\varepsilon_{\min}=\sigma r$. In this way one has for the upper
diagonal element in (\ref{28}). \be \Delta \Lambda (+,+)
=\frac{2\sigma}{\pi^2 r} \delta (1-\cos \theta_{\ver \ver'})
 \int^{\mu/\sigma r}_1 d\tau \frac{\tau+1}{\sqrt{\tau^2-1}}
 \cos (a\sqrt{\tau^2-1}) \theta (\mu-\sigma r)\label{29}\ee
 with $a=\sigma r|r-r'|$, and we keep $r\approx r'$ everywhere except for $a$, since for large
 $a$ (when $r$ is far from $r'$) both $\Lambda$ and  $\Delta \Lambda$ fast decrease.

 In a similar way for the lower diagonal element in (\ref{28}) one has
 \be
 \Delta \Lambda (-,-) =\frac{2\sigma}{\pi^2 r} \delta (1-\cos \theta_{\ver \ver'} )
\int^{\mu/\sigma r}_1 d\tau \frac{\tau-1}{\sqrt{\tau^2-1}}\cos
(a\sqrt{\tau^2-1}) \theta (\mu-\sigma r)\label{30}\ee and taking
$\Lambda_0$ in (\ref{26})  from \cite{12} the resulting form for
$\Lambda$ (\ref{26a}) is
$$
\Lambda( \ver, \ver') \equiv \beta \Lambda_{scalar}+\hat 1 \Lambda_{vector}=
\frac{\beta \sigma}{\pi^2 r}\delta (1-\cos \theta_{\ver \ver'} )
\int_{\mu/\sigma r}^\infty \frac{ d\tau \cos (a\sqrt{\tau^2-1})}{\sqrt{\tau^2-1}} -
$$
\be
-\hat 1 \frac{2\sigma}{\pi^2 r} \delta (1-\cos \theta_{\ver \ver'} )
\int^{\mu/\sigma r}_1  \frac{\tau d\tau}{\sqrt{\tau^2-1}}\cos (a\sqrt{\tau^2-1}).\label{31}\ee

Here $\beta\equiv \gamma_4$, and $\hat 1$ is the unit Dirac matrix, which means,
 that the second term on the  r.h.s. of
(\ref{31}) contributes to the vector part of the resulting mass operator (\ref{21}),
 while the first term contributes to the scalar part. One
should take into account, that in the first term, \be
\Lambda_{scalar} = \frac{\sigma}{\pi^2 r} \delta (1-\cos
\theta_{\ver \ver'} ) \int^\infty_{\tau_{\min}(\mu)}
\frac{d\tau\cos (a\sqrt{\tau^2-1})}{\sqrt{\tau^2-1}} \label{32}\ee
$\tau_{min}(\mu) =\mu/\sigma r$ for $\mu>\sigma r$ and 1
otherwise, so that for large $r,r \gg \frac{\mu}{\sigma}$, one has
the standard $\mu$-independent value \be \Lambda_{scalar} (r\sim
r'> \mu/\sigma) =\frac{\sigma}{\pi^2 r} K_0 (a) \delta (1-\cos
\theta_{\ver\ver'})\label{33}\ee where we have used relations for the McDonald function $K_0$
 \be
K_0(a) =\int^\infty_0 \frac{\cos ax dx}{\sqrt{1+ x^2}}, ~~
\int^\infty_0 da K_0 (a) =\frac{\pi}{2}.\label{34}\ee

One can check that at large $r, r' (r \sim r' > \mu/\sigma)$
 $\Lambda_{scalar} \approx \Lambda_{scalar}^{ (\mu=0)}$ is a smeared $\delta$ -function,
\be \int \Lambda_{scalar}^{(\mu=0)} (\ver, \ver') d^3\ver'
=1.\label{35}\ee However for large $\mu, \mu\gg \sqrt{\sigma}$, $
\Lambda_{scalar}$ is
  different from $\Lambda_{scalar} (\mu=0)$, and
for $r\sim r' < \mu/\sigma$ one has approximately \be
\Lambda_{scalar} (\ver, \ver') =
\Lambda_{scalar}^{(\mu=0)}-f(r,r') \theta(\mu-\sigma r)
\label{36}\ee where \be f(r, r') = \int^{\mu/\sigma r}_1 \frac{
d\tau \cos (a\sqrt{\tau^2-1})}{\sqrt{\tau^2-1}}=\int^{\lambda_0}_1
\frac{d\lambda}{\sqrt{1+\lambda^2}}\cos a\lambda, \label{37}\ee
 and $\lambda_0 ={ \sqrt{\left(\frac{\mu}{\sigma r}\right)^2-1}}$.

Let us now turn to the vector part of interactions, $\Lambda_{vector}$,
$$
\Lambda_{vector} =- \frac{2\sigma}{\pi^2 r} \delta (1-\cos
\theta_{\ver \ver'} ) \int^{\mu/\sigma r}_1  \frac{\tau
d\tau}{\sqrt{\tau^2-1}}\cos (a\sqrt{\tau^2-1}) \theta (\mu-\sigma
r)=
$$
\be -\hat 1\frac{2\sigma}{\pi^2 r} \delta (1-\cos \theta_{\ver
\ver'} )\frac{\sin (a \sqrt{\left(\frac{\mu}{\sigma
r}\right)^2-1})}{a}. \label{38}\ee Returning back to Eq.
(\ref{22}) one can  deduce, that \be M(p_4 =0, \ver, \ver') =
\sqrt{\pi} T_g J^E(\ver, \ver') [\Lambda_{scal} (\ver, \ver')
+\gamma_4 \Lambda_{vector}]\equiv M_{scal}+ \gamma_4
M_{vect.}\label{39}\ee

Taking into account, that at  $r, r'\gg T_g$ and for the Gaussian
$D^E(x)$ one has from (\ref{7}) \be J^E(\ver,\ver')
\cong\frac{(\ver\ver')}{r r'} 2 T_g \sqrt{\pi} D(0) \min (r,
r')\label{40}\ee one has for $M$ at $r, r'\gg T_g$ and for
 $r\cong r'$ \be M_{scal} (r,r') =\sigma r (\tilde
\delta^{(3)} (\ver, \ver') -\xi (\ver, \ver'))\label{41}\ee where
$\tilde \delta^{(3)} (\ver, \ver') \equiv\Lambda^{(\mu=0)}_{scal}
(\ver, \ver')$ and
 \be \xi (\ver, \ver') =\delta_\mu
=\frac{\sigma}{\pi^2 r} (1-\cos \theta_{\ver\ver'}) f (r,
r')\theta (\mu-\sigma r)\label{42}\ee and $f(r, r')$ is defined in
(\ref{37}).

For $M_{vect}$ one has, using (\ref{38}), \be M_{vect} (\ver,
\ver') =-2\sigma r \varphi_\mu (\ver, \ver') \theta(\mu-\sigma
r)\label{43}\ee
where
 \be \varphi_\mu (\ver,
\ver') =\frac{\sigma}{\pi^2 r} \delta (1-\cos \theta_{\ver \ver'}
)\frac{\sin (a \sqrt{\left(\frac{\mu}{\sigma r}\right)^2-1})}{a}.
\label{44}\ee

Note, that one should actually symmetrize all these expressions,
 e.g. $\frac{1}{r} \to \frac{1}{\sqrt{rr'}}$ etc., but we always are in the regime, where $r\approx r'$.
To estimate the magnitude  of nonlocal kernel $M_{scal} (\ver, \ver')$ and $M_{vect} (\ver, \ver')$ it
is convenient to introduce in (\ref{18}) the local limit of the mass operator, namely
\be \bar M_{scal, vect} (r)=  \int d^3 \ver'
M_{scal, vect}  (\ver, \ver').\label{45}\ee

Exploiting the equalities
\be \int d^3 \ver'\tilde \delta^{(3)}
 (\ver, \ver')=\int d^3 \ver'
\xi  (\ver, \ver')
=\int d^3 \ver'
\varphi_\mu(\ver, \ver')=1\label{46}\ee
one arrives at the expressions
\be
\bar M_{scal}(r) =\sigma r \theta(\sigma r-\mu),~~ \bar M_{vect} = - 2 \sigma r \theta (\mu-\sigma r).\label{47}\ee

In the next section we shall discuss approximations made in deriving (\ref{41}), (\ref{43}), (\ref{47})
and physical implications of these results.

\section{ Discussion of results}

Results of the previous section Eqs. (\ref{41}), (\ref{43}), (\ref{47}), can be formulated as follows.
 The relativistic WKB analysis leads to the $\mu$-dependent modification of the confining string, where
  the piece $[0,\mu/\sigma]$ of the string near the
  origin of the string (situated at the  heavy quark position in case of heavy-light quark, or at the string
  junction position in the case of baryons), is dissolved, and the  linear confinement starts beyond the critical
   radius $r_{cr} =\mu/\sigma$. Moreover, an attractive vector interaction appears in the same interval with
    the
   average magnitude $\lan \bar M_{vect}\ran \sim \mu$.

   These conclusions should be taken as qualitative.
First of all, the WKB method is not a good approximation  at small distances, and  we have omitted
 exponentially damped part of $\psi_n(\vez)$ in  the spectrum,
  therefore the inner part of the string is to some extent delocalized (see \cite{12}
 for details) and smoothed.

Secondly, we have not taken into account a possible modification and destruction of the vacuum
due to  the influence of high density quark matter, which might
decrease $\sigma$ or cancel the string completely (as it is happens
 in the thermal phase transition \cite{6}).
The phenomenon of this kind was observed  on the lattice \cite{14} where deconfinement
temperature decreased under the influence of applied external Abelian field.

If however, no density
induced vacuum deconstruction
takes place, then the resulting physical picture according to Eqs. (\ref{47}), is
the net decreasing of confinement in the inner region of some ensemble of quarks, and appearance
of attractive vector potential of the order of $\mu$ acting on each quark.
 This may cause creation of deconfined
bubbles consisting of $3n$ quarks, $n=2,3,..$ in the midst of the nuclear
medium, and dynamically is similar to the $3n q$
bag   formation, which  was studied  before in the framework of the
Quark Compound Bag model \cite{15}.  Note, however, that bag boundary conditions might be strongly
 modified as compared to the standard MIT bag model.
A quantitative analysis of this situation needs a more accurate analysis of
 the $3n$ quark system using nonlinear equations for the $3nq$
 Green's function, generalizing Eq. (\ref{14}).

 The formation of these high-density $3nq$ bubbles may be connected with the explanation
 so-called cumulative effects in the hadron-nucleus(and nucleus-nucleus) collisions, for an example of this discussion
see \cite{16} and refs. therein.

The author is grateful to N.O.Agasian and A.B.Kaidalov for useful discussions.

 The work  is supported
  by the Federal Program of the Russian Ministry of Industry, Science, and Technology
  No.40.052.1.1.1112,  by the
grant  of RFBR No. 06-02-17012, and by the grant for scientific schools NSH-843.2006.2.

\end{document}